\documentclass[aps,prl,reprint,preprintnumbers,showpacs,floatfix,nofootinbib,superscript address,longbibliography]{revtex4-2}
\usepackage[utf8]{inputenc}
\usepackage{amssymb}
\usepackage{hhline}
\usepackage{amsmath}
\usepackage{mathtools}
\usepackage[dvipsnames]{xcolor}
\usepackage{multirow,tabularx}
\usepackage{graphicx}
\usepackage{natbib}

\usepackage{xspace}
\usepackage{xstring}
\usepackage{titlesec}
\usepackage{parskip}
\allowdisplaybreaks
\usepackage{braket}
\usepackage[stable]{footmisc}

\newcommand*\diff{\mathrm{d}}
\newrobustcmd{\pea}[1]{%
	\emph{#1}\textbf{\ \ \ ---}
}
\titleformat{\paragraph}[runin]{\normalfont\normalsize\bfseries}{\emph\theparagraph}{1em}{\pea}

\newcommand*{\ie}{i.e.\@\xspace}
\newcommand*{\eg}{e.g.\@\xspace}
\newcommand*{\cf}{c.f.\@\xspace}
\newcommand*{\fig}{Fig.\@\xspace}
\newcommand*{\eq}{Eq.\@\xspace}
\newcommand*{\tab}{Tab.\@\xspace}
\newcommand*{\eqs}{Eqs.\@\xspace}
\newcommand*{\wrt}{w.r.t.\@\xspace}
\newcommand*{\rhs}{r.h.s.\@\xspace}

\newcommand*{\ex}{\mathrm{e}}

\usepackage{hyperref}
\hypersetup{%
     colorlinks = true,%
     linkcolor = Blue,%
     citecolor = Blue,%
     filecolor = Blue,%
     urlcolor = Blue%
     }%

\begin{document}

\title{On Non-Minimal Couplings to Gravity and Axion Isocurvature Bounds}

\author{Claire Rigouzzo}
\email{claire.rigouzzo@kcl.ac.uk}
\affiliation{Laboratory for Theoretical Particle Physics and Cosmology,\\
	King's College London, London, United Kingdom}
\author{Sebastian Zell}
\email{sebastian.zell@lmu.de}
\affiliation{Arnold Sommerfeld Center, Ludwig-Maximilians-Universit\"at, Theresienstraße 37, 80333 M\"unchen, Germany}
\affiliation{Max-Planck-Institut für Physik, Boltzmannstr. 8, 85748 Garching b.\ M\"unchen, Germany}


\begin{abstract}
	For axions present during inflation, it has been shown that a non-minimal coupling $\xi_\sigma$ of the inflaton to gravity worsens isocurvature bounds \cite{Rigouzzo:2025hza}, while a non-minimal coupling $\xi_\rho$ of the radial Peccei-Quinn field can alleviate them \cite{Graham:2025iwx}. We analyze the simultaneous presence of both couplings and determine when one effect dominates the other, in both the metric and Palatini formulations of gravity. The two tendencies interpolate smoothly, but introducing a non-minimal inflaton coupling reduces the viable interval of $\xi_\rho$ in which isocurvature bounds can be alleviated while avoiding backreaction on the inflationary dynamics.
	We illustrate our findings in Palatini Higgs inflation and Starobinsky inflation. 
\end{abstract}

\maketitle

\section{Introduction}
Axions \cite{Peccei:1977hh,Weinberg:1977ma,Wilczek:1977pj} are among the best-motivated proposals for physics beyond the Standard Model. They provide a viable dark matter candidate and, moreover, offer a solution to the strong CP problem\footnote
{There are indications that quantum gravity does not tolerate eternal de Sitter states due to a fundamental inconsistency caused by quantum breaking \cite{Dvali:2013eja,Dvali:2014gua,Dvali:2017eba}. This turns the existence of an axion from a naturalness question into a consistency requirement \cite{Dvali:2018dce,Dvali:2022fdv}.}
(see \eg the review \cite{DiLuzio:2020wdo}). In the original proposal \cite{Peccei:1977hh}, axions arise as angular Goldstone bosons associated with the breaking of a global Peccei-Quinn (PQ) symmetry. While other ultraviolet completions have been proposed -- including constructions based on extra dimensions \cite{Witten:1984dg} (see also \cite{Reece:2024wrn}), local gauge invariance \cite{Dvali:2005an,Dvali:2013cpa,Dvali:2017mpy,Dvali:2022fdv}, and Einstein-Cartan gravity \cite{Mercuri:2009zi,Karananas:2024xja,Karananas:2025ews} -- in this work we focus on PQ axions.

The cosmological evolution of axions depends on whether the PQ symmetry is broken before or after inflation. If the symmetry is broken after inflation, topological defects form and may overclose the universe in certain scenarios \cite{Sikivie:1982qv}. If PQ symmetry is broken before inflation (and is not restored afterwards), the presence of a massless axion field during inflation induces isocurvature fluctuations \cite{Turner:1990uz}. Their non-observation, especially in cosmic microwave background (CMB) measurements \cite{Planck:2018jri}, imposes stringent constraints on the compatibility of pre-inflationary axions with inflation. Avoiding these isocurvature bounds generally requires a large axion decay constant $f_a$ or a low inflationary Hubble scale $H$.

Some of the leading inflationary models favored by current CMB data \cite{Planck:2018jri, BICEP:2021xfz} -- such as Starobinsky inflation \cite{Starobinsky:1980te}, the metric \cite{Bezrukov:2007ep} and Palatini \cite{Bauer:2008zj} versions of Higgs inflation, and some classes of attractor models \cite{Kallosh:2013tua, Galante:2014ifa}\footnote
{The main feature of $\alpha$-attractor models, a pole in the non-canonical kinetic term, can be obtained from a negative non-minimal coupling, $\xi_\rho <0$ \cite{Kallosh:2013hoa, Kallosh:2013maa}. In this case, it can be possible to alleviate isocurvature bounds \cite{Rigouzzo:2025hza}. However, $\alpha$-attractor models can also be derived independently of a non-minimal coupling \cite{Kallosh:2013hoa,Kallosh:2013yoa}.}
 -- share a common structural feature: they all involve a non-minimal coupling of the inflaton $\sigma$ to gravity of the form $\xi_\sigma \sigma^2 R$, where $R$ is the Ricci scalar and $\xi_\sigma\gg1$ a corresponding coupling constant. Recently, however, we pointed out that such a non-minimal coupling inevitably reduces the inflationary value $f_a^{\text{(inf)}}$ of the decay constant \cite{Rigouzzo:2025hza},
\begin{equation} \label{faInfSimple}
	f_a^{\text{(inf)}} = \frac{f_a}{\sqrt{1 + \xi_\sigma \frac{\sigma^2}{M_P^2}}} \;,
\end{equation}
where $M_P$ is the Planck mass. 

It has long been known \cite{Linde:1991km,Higaki:2014ooa,Choi:2014uaa,Chun:2014xva,Fairbairn:2014zta,Ballesteros:2016xej} that a reduced value $f_a^{\text{(inf)}} < f_a$ strengthens isocurvature constraints, while only $f_a^{\text{(inf)}} > f_a$ can relax them.\footnote
{See \cite{Dvali:1995ce,Dine:2004cq,Jeong:2013xta,Higaki:2014ooa,Kawasaki:2014una,Nakayama:2015pba,Takahashi:2015waa,Kawasaki:2015lpf,Nomura:2015xil,Barenboim:2024akt,Berbig:2024ufe} for further approaches to evading isocurvature bounds.}
Contrary to earlier claims \cite{Tenkanen:2019xzn}, this implies that Palatini Higgs inflation is incompatible with isocurvature bounds \cite{Rigouzzo:2025hza}, while for Starobinsky inflation and metric Higgs inflation a comparable tension already exists even without considering the effect of the modified decay constant.

On the other hand, one may also introduce a non-minimal coupling $\xi_\rho \rho^2 R$ for the radial component $\rho$ of the PQ field \cite{Vilenkin:1982wt,Kofman:1985zx,Linde:1990yj}, with coupling constant $\xi_\rho$. It was recently shown that such a coupling can increase $f_a^{\text{(inf)}}$ and thereby relax isocurvature bounds \cite{Graham:2025iwx}.\footnote
{Furthermore, a non-minimal coupling $\xi_\rho$ can be used to drive inflation with the radial PQ-field \cite{Fairbairn:2014zta,Ballesteros:2016xej,Boucenna:2017fna,Ballesteros:2021bee} (see also \cite{Hashimoto:2021xgu,Barenboim:2024akt,Barenboim:2024xxa}).}
In the relevant parameter regime, the inflationary decay constant takes the approximate form \cite{Graham:2025iwx}
\begin{equation} \label{previousResult}
	f_a^{\text{(inf)}} \approx \sqrt{\frac{12\xi_\rho}{\lambda_\rho}}\, H \;,
\end{equation}
where $\lambda_\rho$ is the self-coupling of $\rho$. A sufficiently large $\xi_\rho$ and sufficiently small $\lambda_\rho$ can therefore yield $f_a^{\text{(inf)}} > f_a$, alleviating isocurvature constraints.

In this paper, we \emph{simultaneously} consider the inflaton coupling $\xi_\sigma$ and the PQ-field coupling $\xi_\rho$, thus combining the strengthening of isocurvature bounds induced by $\xi_\sigma$ \cite{Rigouzzo:2025hza} with the relaxation enabled by $\xi_\rho$ \cite{Graham:2025iwx}. In brief, we find that these effects interpolate smoothly: For sufficiently small $\xi_\rho$, the non-minimal inflaton coupling worsens isocurvature bounds as in \eqref{faInfSimple}, whereas for sufficiently large $\xi_\rho$, the inflationary decay constant becomes insensitive to $\xi_\sigma$ and the result \eqref{previousResult} is recovered. However, the parameter space in which \eqref{previousResult} applies depends on $\xi_\sigma$. While the lower bound on $\xi_\rho$ above which isocurvature constraints can be alleviated is essentially independent of $\xi_\sigma$, a non-minimal inflaton coupling reduces the upper bound on $\xi_\rho$. This limits the maximal enhancement of the inflationary decay constant. As we shall demonstrate, the restriction on $\xi_\rho$ arises primarily from the requirement that the non-minimally coupled PQ field must not interfere with the dynamics of inflation.

Whereas the result of \cite{Graham:2025iwx} was obtained in the metric formulation of General Relativity (GR), we extend the analysis to Palatini gravity, showing that the choice of formulation does not significantly affect the outcome. For each formulation, we select a representative inflationary model for detailed study: Palatini Higgs inflation \cite{Bezrukov:2007ep,Bauer:2008zj}, following \cite{Rigouzzo:2025hza}, and the Starobinsky model \cite{Starobinsky:1980te}. Besides warm inflation (see \cite{Berera:1995ie} and \cite{McLerran:1990de,Berghaus:2019whh,Laine:2021ego,Mirbabayi:2022cbt,Berghaus:2025dqi}), \cite{Graham:2025iwx} also considered Starobinsky inflation, though without accounting for the effect of $\xi_\sigma$ on $f_a^{\text{(inf)}}$. We find that the mechanism proposed in \cite{Graham:2025iwx} for alleviating isocurvature constraints indeed applies to Starobinsky inflation, but only within a smaller region of parameter space than was considered in \cite{Graham:2025iwx}.

The paper is organized as follows. In section \ref{sec:model}, we introduce the model featuring both the axion and the inflaton coupled non-minimally to gravity. We show that the decay constant generically depends on the non-minimal couplings and discuss the resulting impact on isocurvature constraints. In section \ref{sec:non_minimal_coupling}, we analyze the limiting case where only the axion couples non-minimally to gravity, \ie $\xi_\sigma=0$. Finally, section \ref{sec:full} presents the complete analysis including both non-minimal couplings, in both the Palatini and metric formulations of GR. As a key example, we demonstrate that it is possible to satisfy isocurvature bounds in Starobinsky inflation, albeit within a restricted region of parameter space.

\paragraph{Convention:} We use the metric signature $(-1, 1, 1, 1)$.

\section{The model}
\label{sec:model}
We consider an inflaton field $\sigma$ with non-minimal coupling $\xi_\sigma$ and the PQ-field $\Phi$ with non-minimal coupling $\xi_\rho$:\footnote
{Note that these terms are allowed by symmetry, and of mass dimension less or equal to $4$, so one should generically include them.}
\begin{align} \label{actionStart}
	\mathcal{L} & =   \left(\frac{M_P^2}{2}+ \xi_\rho |\Phi|^2 + \frac{1}{2} \xi_\sigma \sigma^2 \right) R \nonumber\\
	&  - \partial_\mu \Phi \partial^\mu  \Phi^\star - V_\rho - \frac{1}{2} \partial_\mu \sigma \partial^\mu  \sigma - V_\sigma \;, 
	\end{align}
	where we assume $\xi_\sigma>0$, $\xi_\rho>0$ and
	\begin{equation}
	  V_\rho = \lambda_\rho \left(|\Phi|^2 - \frac{1}{2} f_a^2\right)^2 \label{potential} \;.
\end{equation} 
Splitting the PQ field into
\begin{equation} \label{splitPhi}
	\Phi = \frac{\rho}{\sqrt{2}} \exp\left(i \varphi\right) \,,
\end{equation}
one recognizes the non-minimal coupling to the scalar curvature:
\begin{equation} \label{omega}
	\Omega^2 = 1 + \frac{\xi_\rho \rho^2}{M_P^2} + \frac{\xi_\sigma \sigma^2}{M_P^2} \;.
\end{equation}
We perform the conformal transformation $g_{\mu\nu} \rightarrow \Omega^{-2} g_{\mu\nu}$, to obtain (see \cite{Rigouzzo:2022yan}) 
\begin{align} 
	\mathcal{L} & =    \frac{M_P^2}{2} R - \frac{\rho^2}{2 \Omega^2} \partial_\mu \varphi \partial^\mu \varphi \nonumber\\ & - \frac{1}{2 \Omega^2}\left(1+ \frac{\zeta 6 \xi_\rho^2 \rho^2}{M_P^2 \Omega^2}\right) \partial_\mu \rho \partial^\mu\rho  - \frac{V_\rho}{\Omega^4} \nonumber\\
	& - \frac{1}{2 \Omega^2}\left(1+ \frac{\zeta 6 \xi_\sigma^2 \sigma^2}{M_P^2 \Omega^2}\right) \partial_\mu \sigma \partial^\mu\sigma  - \frac{V_\sigma}{\Omega^4} \nonumber\\
	& - \frac{1}{\Omega^2} \frac{\zeta 6 \xi_\rho \xi_\sigma \rho \sigma}{M_P^2 \Omega^2} \partial_\mu \rho \partial^\mu\sigma \;. \label{actionEinstein}
\end{align}
As is well-known, the outcome of the conformal transformation depends on the formulation of GR. In order to account for this fact, we introduced the parameter $\zeta$, where $\zeta=1$ for metric GR while $\zeta=0$ in the Palatini case (see \eg \cite{Rigouzzo:2022yan} for more details on the equivalent formulations of GR).

\subsection{Inflationary decay constant} From the first line of \eq \eqref{actionEinstein}, we see that the inflationary value of the decay constant is given by
\begin{equation} \label{faInfPrevious}
	f_a^{\text{(inf)}} = \frac{\rho_{\text{min}}}{\Omega} \;,
	\end{equation}
where $\rho_{\text{min}}$ is the field value that minimizes the potential of field $\rho$ for a given inflationary background.  It is important to note that $f_a^{\text{(inf)}}$ is independent of the non-canonical kinetic term of the radial field $\rho$.\footnote
{This is because we are interested in the minimum of the potential, which happens when the kinetic term vanishes $\partial_\mu \rho=0$.}
As shown in \cite{Rigouzzo:2025hza}, we immediately see why a non-minimal coupling of the inflaton worsens isocurvature bounds: If $\xi_\rho=0$, we get $\rho_{\text{min}}=f_a$ and so $\Omega \gg 1$ decreases the inflationary decay constant \cite{Rigouzzo:2025hza}, as shown in \eq \eqref{faInfSimple}.
 What remains to be done is to evaluate $\rho_{\text{min}}$ for non-vanishing $\xi_\rho$, which we shall do in different scenarios.

\subsection{Isocurvature bounds}
Before that, we will briefly state the known isocurvature bound \cite{Turner:1990uz}.
Assuming that axions make up all of dark matter, it reads \cite{Choi:2014uaa,Chun:2014xva,Fairbairn:2014zta,Ballesteros:2021bee,Rigouzzo:2025hza, Graham:2025iwx}
\begin{equation} \label{isocurvatureBound}
	\frac{H}{2\pi f_a^{\text{(inf)}}} \lesssim 4.6 \cdot  10^{-6} \left(\frac{1.02 \cdot 10^{12} \text{GeV}}{f_a}\right)^{7/12} \;.
\end{equation}
Plugging in the result \eqref{previousResult} of \cite{Graham:2025iwx} for $f_a^{\text{(inf)}}$ then gives
\begin{equation} \label{isocurvatureBoundLifted}
	\lambda_\rho < 10^{-8} \xi_\rho \left(\frac{1.02 \cdot 10^{12} \text{GeV}}{f_a}\right)^{7/6} \;.
\end{equation}
Thus, evading isocurvature bounds generically requires a tiny $\lambda_\rho$ or a large $\xi_\rho\gg1$.

\section{Inflaton minimal coupling}
\label{sec:non_minimal_coupling} First, we shall consider the case of a minimally coupled inflaton, $\xi_\sigma=0$, as in \cite{Graham:2025iwx}. The following derivation applies both to the metric and Palatini formulations.
Since there is no kinetic mixing between $\rho$ and $\sigma$, \eq \eqref{actionEinstein} directly gives the effective potential of the two scalar fields:
\begin{equation} \label{UMinimal}
	U = \frac{V_\rho + V_\sigma}{\Omega^4}= \frac{\frac{\lambda_\rho}{4} \left(\rho^2 - f_a^2\right)^2+ V_\sigma}{\left(1 + \frac{\xi_\rho \rho^2}{M_P^2}\right)^2} \;.
\end{equation}
Minimizing $U$, we get
\begin{equation} \label{rhoMinMinimal}
	\rho_{\text{min}} = f_a \frac{\sqrt{1 + \frac{\xi_\rho f_a^2}{M_P^2} + \frac{4\xi_\rho V_\sigma}{\lambda_\rho f_a^2 M_P^2}}}{\sqrt{1 + \frac{\xi_\rho f_a^2}{M_P^2}}} \;,
\end{equation}
and plugging this into \eq \eqref{omega} yields
\begin{equation}
	\Omega^2 = 1 + \frac{\xi_\rho f_a^2}{M_P^2} + \frac{4 \xi_\rho^2 V_\sigma}{\lambda_\rho M_P^4 \left(1 + \frac{\xi_\rho f_a^2}{M_P^2}\right)} \;.
\end{equation}
This determines the inflationary decay constant via \eq \eqref{faInfPrevious}.

As evident from the third line of \eq \eqref{actionEinstein}, the presence of a non-minimal coupling $\xi_\rho$ of the PQ-field leads to a non-trivial coefficient $1/\Omega^2$ of the inflaton kinetic term. In order to avoid changing inflation, we need $\Omega^2\approx 1$. So $\Omega^2-1 \ll 1$ leads to the necessary requirement 
\begin{equation} \label{boundMinimal}
\xi_\rho \ll \min\left(\frac{M_P^2}{f_a^2},\sqrt{\lambda_\rho} \frac{M_P^2}{\sqrt{V_\sigma}}\right)\;,
\end{equation}
where the second part coincides with its counterpart derived in \cite{Graham:2025iwx}.
In general, however, the condition $U \approx V_\sigma$ may not be sufficient for leaving the dynamics of the inflaton unaltered. In order to check this, we can evaluate the correction to the first slow-roll parameter:
	\begin{equation}
		\frac{\sqrt{\epsilon}-\sqrt{\epsilon}\big|_{\xi_\rho=0}}{\sqrt{\epsilon}\big|_{\xi_\rho=0}} \approx \Omega-1 \sim \frac{\xi_\rho \rho^2_\text{min}}{M_P^2} \;,
	\end{equation} 
	where we used that the canonical inflaton field $\chi$ satisfies approximately $\diff \chi \approx \diff \sigma/\Omega$. We reproduce the condition $\Omega-1 \ll 1$, and so for the case of a minimally coupled inflaton, the condition \eqref{boundMinimal} is both sufficient and necessary for not changing $\epsilon$.

If $\xi_\rho$ obeys the bound \eqref{boundMinimal}, plugging \eq \eqref{rhoMinMinimal} into the potential \eqref{UMinimal} shows that $U(\rho_{\text{min}})\approx  V_\sigma$. So $V_\sigma \approx 3 M_P^2 H^2$ and we can approximate
\begin{equation} \label{faInfMinimalApprox}
	f_a^{\text{(inf)}} \approx 	\rho_{\text{min}} \approx  f_a \left(\sqrt{1 + \frac{12\xi_\rho H^2}{\lambda_\rho f_a^2}}\right) \;.
\end{equation}
As it should, the value \eqref{faInfMinimalApprox} of the inflationary decay constant coincides with the result \eqref{previousResult} of \cite{Graham:2025iwx}, which has been derived in the Jordan frame by minimizing	$V_{\text{eff}} \approx V_\rho  - \xi_\rho R |\Phi|^2 \approx V_\rho  - 12 \xi_\rho H^2 |\Phi|^2$. We have rederived this finding in the Einstein frame and shown that it is also applicable in the Palatini formulation of GR.

In summary, the mechanism for lifting isocurvature constraints is effective if \footnote
{\label{fn:minimalUpperBound}Note that $\lambda_\rho f_a^2/H^2 \ll \xi_\rho \ll \sqrt{\lambda_\rho M_P/H}$ implies $\sqrt{\lambda_\rho M_P/H} \gg M_P^2/f_a^2$, and so we only need to consider the second condition of \eq \eqref{boundMinimal}.}
\begin{align}
 &\frac{\lambda_\rho f_a^2}{H^2} \ll \xi_\rho \ll \sqrt{\lambda_\rho} \frac{M_P}{H}\nonumber\\
   \Rightarrow \quad  & f_a^{\text{(inf)}} \approx \sqrt{\frac{12\xi_\rho}{\lambda_\rho}} \, H \, \gtrsim\,  f_a \;, \label{summaryMinimal}
\end{align}
where we took into account \eqs \eqref{boundMinimal} and \eqref{faInfMinimalApprox}.
Such a $\xi_\rho$ only exists if 
\begin{equation} \label{boundInflatonPotential}
\lambda_\rho f_a^4 \ll M_P^2 H^2 \;.
\end{equation}
In general, the maximal temperature during reheating, \ie the transition from inflation to radiation-dominated expansion, fulfills (see \eg \cite{Rubio:2019ypq,Shaposhnikov:2020fdv})
\begin{equation} \label{TMax}
T_{\text{max}} < \left(\frac{90 M_P^2 H^2}{\pi^2 g_\star}\right)^{1/4}  \;.
\end{equation}
This bound is saturated if reheating can be approximated as instantaneous. In this case, the hierarchy \eqref{boundInflatonPotential} and the condition $T_{\text{max}}<f_a$, which is necessary to avoid restoration of the PQ-symmetry, can only be satisfied simultaneously if $\lambda_\rho \ll \left(T_{\text{max}}/f_a\right)^4\ll 1$.
 
\section{General case: full non-minimal couplings}
\label{sec:full}
\subsection{Palatini gravity} Next, we shall include a non-minimal coupling $\xi_\sigma$ of the inflaton, but specialize to the Palatini formulation of GR.  Correspondingly, we set $\zeta=0$ in the Einstein frame action \eqref{actionEinstein}. Since no kinetic mixing of the two scalar fields exits in Palatini GR, the effective potential becomes
\begin{equation}
	U= \frac{V_\sigma+V_\rho}{\Omega^4} = \frac{\frac{\lambda_\rho}{4} \left(\rho^2 - f_a^2\right)^2+ V_\sigma}{\left(1 + \frac{\xi_\rho \rho^2}{M_P^2} + \frac{\xi_\sigma \sigma^2}{M_P^2}\right)^2} \;,
\end{equation}
which is minimized for
\begin{equation}
	\rho_{\text{min}} = f_a \sqrt{\frac{1 + \frac{\xi_\rho f_a^2}{M_P^2} + \frac{\xi_\sigma \sigma^2}{M_P^2}  + \frac{4 \xi_\rho  V_\sigma}{\lambda_\rho f_a^2 M_P^2}}{1 + \frac{\xi_\rho f_a^2}{M_P^2} + \frac{\xi_\sigma \sigma^2}{M_P^2}}} \;,
\end{equation}
and so
\begin{equation}
	\Omega^2 = \frac{\left(1 + \frac{\xi_\rho f_a^2}{M_P^2} + \frac{\xi_\sigma \sigma^2}{M_P^2}\right)^2 + \frac{4 \xi_\rho^2 V_\sigma}{\lambda_\rho M_P^4}}{1 + \frac{\xi_\rho f_a^2}{M_P^2} + \frac{\xi_\sigma \sigma^2}{M_P^2}} \;.
\end{equation}
As before, this determines the inflationary decay constant via \eq \eqref{faInfPrevious}.

We can now split the full non-minimal coupling $\Omega$ into the axion and inflaton dependent parts. Introducing the notation
\begin{equation}
	\Omega_0^2 \equiv 1 + \frac{\xi_\sigma \sigma^2}{M_P^2} \;,
\end{equation}
we shall assume that (already in the absence of an axion) $\Omega_0^2\gg 1$. Thus, a necessary condition for avoiding backreaction on the inflaton is to require $\Omega^2 \approx \Omega_0^2$, and so $\Omega^2 - \Omega_0^2\ll \Omega_0^2$ implies
\begin{equation}
	\label{boundNonMinimal}
	\xi_\rho \ll \min\left(\Omega_0^2 \frac{M_P^2}{f_a^2}, \sqrt{\lambda_\rho} \frac{M_P}{H}\right) \;,
\end{equation}
where we self-consistently used that $3 M_P^2 H^2 \approx V_\sigma/\Omega_0^4 \approx M_P^4 V_\sigma/(\xi_\sigma^2 \sigma^4)$. The first condition of this bound is less stringent than its counterpart \eqref{boundMinimal} in the minimally coupled case. Provided \eq \eqref{boundNonMinimal} is satisfied, we can approximate
\begin{equation} 
	f_a^{\text{(inf)}}  \approx  f_a \sqrt{\frac{1}{\Omega_0^2} + \frac{12\xi_\rho H^2}{\lambda_\rho f_a^2}}  \;, \label{faInfNonMinimal}
\end{equation}
with the limits 
\begin{equation}
	f_a^{\text{(inf)}} \approx \begin{cases} \displaystyle
		\frac{f_a}{\Omega_0} \quad &\text{for} \quad  \xi_\rho \ll \frac{\lambda_\rho f_a^2}{H^2 \Omega_0^2} \\ \\
		\displaystyle \sqrt{\frac{12\xi_\rho }{\lambda_\rho}}H \quad &\text{for} \quad  \xi_\rho \gg \frac{\lambda_\rho f_a^2}{H^2 \Omega_0^2} \;.
	\end{cases}
\end{equation}
The first line coincides with the result \eqref{faInfPrevious} and so leads to $f_a^{\text{(inf)}}\ll f_a$, which results in a worsening of isocurvature bounds. Interestingly, the second line is identical to the finding obtained for a minimally coupled inflaton (see \eqs \eqref{previousResult} and \eqref{summaryMinimal}). Thus, if the coupling to the axion is sufficiently large, a non-minimal coupling of the inflaton no longer affects the inflationary decay constant. However, comparison of \eqs \eqref{faInfMinimalApprox} and \eqref{faInfNonMinimal} shows that for a given choice of parameters, $f_a^{\text{(inf)}}$ is always smaller for a non-minimally coupled inflaton as compared to the minimally coupled case.

In summary, the relevant parameter space for lifting isocurvature constraints is\footnote
{For an analogous reason as in footnote \ref{fn:minimalUpperBound}, we can drop $\Omega_0^2 M_P^2/f_a^2$ in comparison to $\sqrt{\lambda_\rho}M_P/H$.}
\begin{align}
	&\frac{\lambda_\rho f_a^2}{H^2} \ll \xi_\rho \ll \sqrt{\lambda_\rho} \frac{M_P}{H} \nonumber\\
	\Rightarrow \quad &  f_a^{\text{(inf)}} \approx \sqrt{\frac{12\xi_\rho}{\lambda_\rho}}\, H \, \gtrsim\,  f_a \;. \label{summaryNonminimal}
\end{align}
It is important to note that $f_a^{\text{(inf)}} \approx \sqrt{12\xi_\rho/\lambda_\rho} H$ already holds for smaller $\xi_\rho \gg \lambda_\rho f_a^2/(\Omega_0^2 H^2)$, but achieving $f_a^{\text{(inf)}}>f_a$ requires the more stringent lower bound $\xi_\rho\gg \lambda_\rho f_a^2/H^2$. Thus, \eq \eqref{summaryNonminimal} coincides with its counterpart \eqref{summaryMinimal} of a minimally coupled inflaton.
 However, the bounds on $\xi_\rho$ of \eq \eqref{summaryMinimal} are only necessary conditions for the validity of lifting isocurvature constraints, but in general they are not sufficient. In particular, the upper bound on $\xi_\rho$ can be significantly stronger than shown in \eq \eqref{summaryNonminimal}, as we shall demonstrate shortly (see \eq \eqref{boundPalatini}).

\paragraph*{Palatini Higgs inflation} We can come back to isocurvature bounds in Palatini Higgs inflation \cite{Bezrukov:2007ep, Bauer:2008zj}, as studied in \cite{Rigouzzo:2025hza}. Then $\sigma$ is the Higgs field (in unitary gauge), and so $V=\lambda/4 \,\sigma^4$, where $\lambda$ is the high-energy value of the Higgs self-coupling. In this concrete model, we can evaluate the first slow-roll parameter and its leading correction:
\begin{equation}
	\frac{\sqrt{\epsilon}-\sqrt{\epsilon}\big|_{\xi_\rho=0}}{\sqrt{\epsilon}\big|_{\xi_\rho=0}}  \sim  \Omega-1 \sim \xi_\rho \frac{\rho_{\text{min}}^2}{M_P^2} \,.
\end{equation}
Requiring it to be small implies (\cf \eq \eqref{faInfNonMinimal})
\begin{equation}
	\label{boundPalatini}
	\xi_\rho \ll \min\left(\frac{M_P^2}{f_a^2}, \frac{\sqrt{\lambda_\rho}}{\Omega_0} \frac{M_P}{H}\right) \;.
\end{equation}
This condition, derived from the first derivative of the potential, is significantly more restricting than the condition \eqref{boundNonMinimal} derived from the value of the potential itself. Comparing \eq \eqref{boundPalatini} with its counterpart \eqref{boundMinimal} in the minimally coupled case, we see that the first conditions coincide while the second one is considerably stronger in the non-minimally coupled case. 

\begin{figure*}
	\centering
	\includegraphics[width=0.9\textwidth]{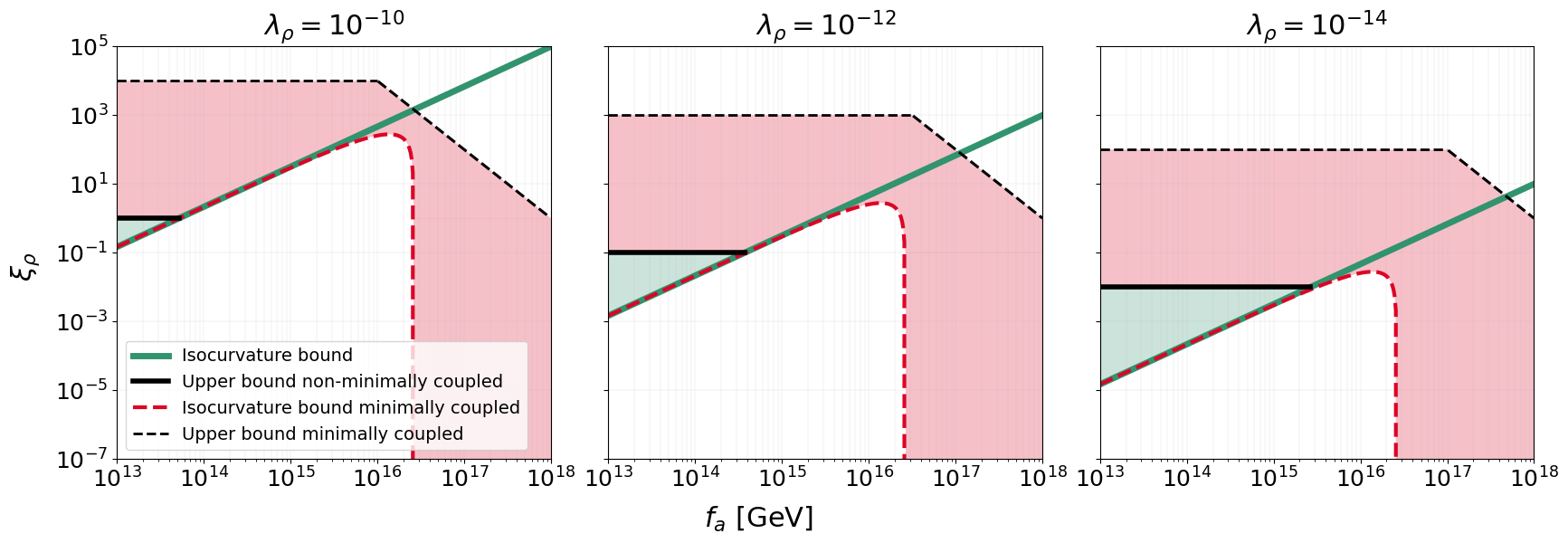}
	\caption{Constraints on $\xi_\rho$ in Palatini Higgs inflation as a function of $f_a$, for $H=10^{9}$ GeV and some choices of $\lambda_\rho$. The green line represents the lower bound on $\xi_\rho$ for which the inflationary decay constant \eqref{faInfNonMinimal} fulfills the isocurvature bound \eqref{isocurvatureBound}.
		The red curve shows the would-be lower bound on $\xi_\rho$ if the inflaton were minimally coupled, which is found by plugging
		\eq \eqref{faInfMinimalApprox} into \eq \eqref{isocurvatureBound}. Furthermore, the solid black line represents the upper bound on $\xi_\rho$ from imposing the non-backreaction condition of \eq \eqref{boundPalatini} (with $\Omega_0 \sim 2 \cdot10^4$) and the dashed black line corresponds to the would-be upper bound for the minimally coupled case. In the green region, isocurvature bounds are obeyed and the red region would only be viable if the influence of the inflaton non-minimal coupling on axions were neglected. That the red region on the right extends down to vanishing $\xi_\rho$ reflects the fact that Palatini Higgs inflation would obey isocurvature bound if the effect of the inflaton non-minimal coupling were not taken into account \cite{Tenkanen:2019xzn}.}
	\label{fig:PalatiniHiggs}
\end{figure*}

Matching the amplitude of CMB perturbations requires $\xi_\sigma\sim 10^{7}$ and during inflation $\sigma \sim \sqrt{N} M_P$ with CMB generation at $N\approx 51$ (see \cite{Shaposhnikov:2020fdv}). Consequently, we have $\Omega_0^2 \approx 5\cdot 10^8$ and moreover $H\approx 10^{-2} M_P/\xi_\sigma\approx 10^{-9} M_P$. In general, the second condition in  \eq \eqref{boundPalatini} is more stringent and the largest admissible non-minimal coupling becomes $\xi_{\rho, \, \text{max}} \sim 10^5 \sqrt{\lambda_\rho}$. Plugging this into the bound \eqref{isocurvatureBoundLifted} gives 
\begin{equation} \label{boundLambdaPalatiniHI}
	\lambda_\rho < 10^{-6} \left(\frac{10^{12} \text{GeV}}{f_a}\right)^{7/3} \;.
\end{equation}  
On the other hand, reheating in Palatini Higgs inflation can be well approximated as instantaneous \cite{Rubio:2019ypq,Dux:2022kuk}, and so \eq \eqref{TMax} gives $T_{\text{max}}\approx  4 \cdot 10^{13} \text{GeV}$,
where we took $g_\star\approx 100$. Since $\Omega_0$ changes rapidly during reheating \cite{Rubio:2019ypq, Dux:2022kuk}, the relevant value the axionic decay constant and the precise condition for non-restoration of PQ-symmetry remain to be determined. Nevertheless, it is reasonable to expect that $T_{\text{max}}>f_a$ will excite the radial mode of the PQ-field (see also \cite{Graham:2025iwx}). Thus, plugging $f_a>4 \cdot 10^{13} \text{GeV}$ in the bound \eqref{boundLambdaPalatiniHI} shows that isocurvature bound can only be lifted at the price of an extremely small $\lambda_\rho \lesssim 10^{-9}$. For different values of $\lambda_\rho$, we show in \fig \ref{fig:PalatiniHiggs} the influence of the inflationary decay constant \eqref{faInfNonMinimal} on isocurvature bounds.

\subsection{Metric gravity}
For metric GR, the situation is more complicated due to the kinetic mixing in the last line of \eq \eqref{actionEinstein}. In order to obtain two at least approximately independent scalar fields, one needs to perform an appropriate shift of $\sigma$,
\begin{equation} \label{shiftMetric}
	\sigma \rightarrow \sigma + f(\rho,\sigma) \;,
\end{equation} 
where $f(\rho,\sigma)$ depends on both  $\rho$ and $\sigma$. This generates additional terms in the potential of $\rho$, which depend on $V_\sigma$. As a result, $\rho_{\text{min}}$ might change, although this has to be analyzed on a case by case basis. 

\paragraph*{Starobinsky inflation} We shall work out one particularly important model of a non-minimally coupled inflaton in metric GR: Starobinsky inflation \cite{Starobinsky:1980te}. The theory is
\begin{align}
		\mathcal{L}&= \left(\frac{M_P^2}{2}+ \xi_\rho |\Phi|^2 \right) R  + \frac{M_P^2}{12 M^2} R^2\nonumber\\
		&  - \partial_\mu \Phi \partial^\mu  \Phi^\star - V_\rho  \;, \label{actionStartStarobinsky}
\end{align}
which replaces \eq \eqref{actionStart}. As usual, one can then introduce an auxiliary scalar field $\sigma$ to replace the $R^2$ term: 
\begin{align}
		\mathcal{L} &= \left(\frac{M_P^2}{2}+ \frac{\xi_\rho \rho^2}{2} + \frac{M_P^2 \sigma^2}{6 M^2}   \right) R - \frac{M_P^2}{12 M^2} \sigma^4 
		 \nonumber\\
		 & - \frac{1}{2} \partial_\mu \rho \partial^\mu \rho -\frac{\rho^2}{2} \partial_\mu \varphi \partial^\mu \varphi - V_\rho  \;, \label{actionStartStarobinsky2}
\end{align}
where we also plugged in the decomposition \eqref{splitPhi} of $\Phi$.
As before, its presence should not alter inflationary dynamics, which implies
\begin{equation}\label{boundXiStarobinsky}
	\xi_\rho \ll \frac{M_P^2 \sigma^2}{M^2 \rho_{\text{min}}^2} \;.
\end{equation}
What is unique about Starobinsky inflation is that the action \eqref{actionStartStarobinsky2} does not contain a kinetic term for $\sigma$, \ie it is only generated through the conformal transformation. Therefore, it is convenient to perform the shift \eqref{shiftMetric} already in \eq \eqref{actionStartStarobinsky2}. Redefining
\begin{equation} \label{fieldRedefinitionStarobinsky}
	\sigma^2 \rightarrow \sigma^2 - \frac{3 \xi_\rho \rho^2  M^2}{M_P^2} \;,
\end{equation}
we get\footnote
{Note that unlike in the previously considered inflationary scenarios, imposing $\Omega^2-1 \ll 1$ does not lead to an upper bound on $\xi_\rho$ in terms of $M_P^2/f_a^2$ (\cf \eqs \eqref{summaryMinimal} and \eqref{summaryNonminimal}) since the non-minimally coupling $\rho^2 R$ can be removed by the redefinition \eqref{fieldRedefinitionStarobinsky} of the non-propagating field $\sigma$.}
\begin{align}
	\mathcal{L} &= \left(\frac{M_P^2}{2} + \frac{\sigma^2 M_P^2}{6 M^2}   \right) R - \frac{M_P^2}{12 M^2} \sigma^4 
	\nonumber\\
	& - \frac{1}{2} \partial_\mu \rho \partial^\mu \rho -\frac{\rho^2}{2} \partial_\mu \varphi \partial^\mu \varphi \nonumber\\
	& - V_\rho + \frac{\xi_\rho}{2}\rho^2 \sigma^2 - \frac{3\xi_\rho^2 M^2}{4 M_P^2}\rho^4 \;. \label{actionStarobinskyRedefined}
\end{align}
The first line of \eq \eqref{actionStarobinskyRedefined} describes pure Starobinsky inflation in the absence of an axion. Therefore, we can perform the conformal transformation $g_{\mu\nu} \rightarrow \Omega^{-2} g_{\mu\nu}$ with $\Omega^2 =1 + \sigma^2/(3 M^2)$ to obtain
\begin{align}
	\mathcal{L} &= \frac{M_P^2}{2} R - \frac{M_P^2}{12 M^2} \frac{\sigma^4}{\Omega^4} - \frac{M_P^2 \sigma^2}{3 M^4 \Omega^4} \partial_\mu \sigma \partial^\mu \sigma
	\nonumber\\
	& - \frac{1}{2 \Omega^2}\partial_\mu \rho \partial^\mu \rho -\frac{\rho^2}{2 \Omega^2} \partial_\mu \varphi \partial^\mu \varphi \nonumber\\
	& -  \frac{1}{\Omega^4}\left(V_\rho - \frac{\xi_\rho}{2}\rho^2 \sigma^2 + \frac{3\xi_\rho^2 M^2}{4 M_P^2}\rho^4\right) \;.
\end{align}
Introducing the canonically normalized inflaton $\chi$ via
\begin{equation}
	\sigma^2 = 3M^2 \left(\ex^{\sqrt{2/3} \chi / M_P} - 1 \right) \;,
\end{equation}
and using the potential term as defined in  \eq \eqref{potential},
we arrive at 
\begin{align}
		\mathcal{L} & = \frac{M_P^2}{2} R - \frac{3}{4} M_P^2 M^2 \left( 1 - \ex^{-\sqrt{2/3} \chi / M_P} \right)^2  \nonumber\\
		&- \frac{1}{2}\partial_\mu \chi \partial^\mu \chi -\frac{1}{2 \ex^{\sqrt{2/3}\chi/M_P}} \partial_\mu \rho \partial^\mu \rho  \nonumber\\& -\frac{1}{2 \ex^{\sqrt{2/3}\chi/M_P}}\rho^2 \partial_\mu \varphi \partial^\mu \varphi \nonumber\\& - \frac{\lambda_\rho}{4 \ex^{2 \sqrt{2/3}\chi/M_P}} \Bigg(\left(1+\frac{3 \xi_\rho^2 M^2}{ \lambda_\rho M_P^2}\right)\rho^4 \nonumber\\& - \left(\frac{6 \xi_\rho M^2}{\lambda_\rho}\left(\ex^{\sqrt{2/3}\chi/M_P}-1\right)+2 f_a^2\right)\rho^2 +f_a^4\Bigg) \;. \label{actionStarobinskyFinal}
\end{align}

Therefore, we can approximate the potential of $\rho$ during inflation as
\begin{align}
		U_\rho &\approx \frac{9\lambda_\rho}{64 N^2} \Bigg(\left(1+\frac{12 \xi_\rho^2 H^2}{\lambda_\rho M_P^2}\right)\rho^4 \nonumber\\& - 2\left(f_a^2 +  \frac{16 N \xi_\rho H^2}{\lambda_\rho}\right)\rho^2 +f_a^4\Bigg)\;, \label{VRhoStarobinsky}
\end{align}
where we used that $H=\sqrt{V/3}/M_P\approx M/2$ and $\chi \sim M_P \sqrt{3/2}\ln(4N/3)$ in Starobinsky inflation (see \cite{Rigouzzo:2025hza}). Thus, the minimum of the potential is at
\begin{equation} 
	\rho_{\text{min}} \approx f_a \sqrt{\frac{1 +  \frac{16 N \xi_\rho H^2}{\lambda_\rho f_a^2}}{1+\frac{12 \xi_\rho^2 H^2}{\lambda_\rho M_P^2}}} \;,
\end{equation}
and we arrive at the inflationary decay constant
\begin{equation} \label{fInfStarobinskyNonminimal}
	f_a^{\text{(inf)}} = \frac{\rho_{\text{min}}}{\Omega} \approx f_a  \sqrt{\frac{\frac{3}{4 N} +  \frac{12 \xi_\rho H^2}{\lambda_\rho f_a^2}}{1+\frac{12 \xi_\rho^2 H^2 }{\lambda_\rho M_P^2}}} \;.
\end{equation}
We can approximate
\begin{equation} \label{fInfStarobinskyNonminimalApprox}
	f_a^{\text{(inf)}}\simeq \begin{cases}
		\sqrt{\frac{3}{4 N}} f_a \quad &\text{for} \quad \xi_\rho \ll \frac{\lambda_\rho f_a^2}{N H^2}\\
		\sqrt{\frac{12 \xi_\rho H^2}{\lambda_\rho}} \quad &\text{for} \quad    \frac{\lambda_\rho f_a^2}{N H^2} \ll \xi_\rho \ll \sqrt{\lambda_\rho} \frac{M_P}{H}\\
		\frac{M_P}{\sqrt{\xi_\rho}}  \quad &\text{for} \quad \xi_\rho \gg \sqrt{\lambda_\rho} \frac{M_P}{H} 
	\end{cases} \;,
\end{equation}
where we assumed $\sqrt{\lambda_\rho} M_P/H \gg\lambda_\rho f_a^2/(N H^2)$.\footnote
{For large $f_a$ close to $M_P$, it can be possible to achieve a hierarchy $\sqrt{\lambda_\rho} M_P/H \ll \xi_\rho \ll \lambda_\rho f_a^2/(N H^2)$. Then one gets 
	\begin{equation}
		f_a^{\text{(inf)}}\simeq \frac{f_a M_P}{4 \xi_\rho H } \sqrt{\frac{\lambda_\rho}{N}}  \;,
	\end{equation}
	and evidently $f_a^{\text{(inf)}}<f_a$.}
Therefore, if $\xi_\rho$ is too small, we get $f_a^{\text{(inf)}}<f_a$ and isocurvature bounds are strengthened, as derived in \cite{Rigouzzo:2025hza}.

In order to identify viable parts in parameter space, we first need to make sure to fulfill the condition of not altering inflation. From \eq \eqref{actionStarobinskyFinal}, we can read off the leading correction to the first derivative of the potential \wrt $\chi$ (first term in last line of \eq \eqref{actionStarobinskyFinal}): 
\begin{equation}
\frac{\sqrt{\epsilon}-\sqrt{\epsilon}\big|_{\xi_\rho=0}}{\sqrt{\epsilon}\big|_{\xi_\rho=0}}  \sim  \xi_\rho \frac{\rho_{\text{min}}^2}{M_P^2} \;,
\end{equation}
where we assumed $f_a^4 \lambda_\rho \ll M^2 M_P^2 N$.
This leads to the condition
\begin{equation} \label{boundXiStarobinskyExact}
	 \xi_\rho \frac{\rho_{\text{min}}^2}{M_P^2} \ll 1 \qquad \Leftrightarrow \qquad \xi_\rho \ll \frac{M_P^2}{N 	f_a^{\text{(inf)\ 2}}}  \;,
\end{equation}
which is stronger than the estimate \eqref{boundXiStarobinsky} since $\sigma^2 \sim N M^2$. We conclude that the third line of \eq \eqref{fInfStarobinskyNonminimalApprox} cannot obey \eq \eqref{boundXiStarobinskyExact}, and in the second line, the admissible interval of $\xi_\rho$ shrinks.

In summary, we can alleviate isocurvature bounds if
\begin{align}
	&\frac{\lambda_\rho f_a^2}{H^2} \ll \xi_\rho \ll \sqrt{\frac{\lambda_\rho}{N}} \frac{M_P}{H} \nonumber\\
	\Rightarrow \quad &  f_a^{\text{(inf)}} \approx \sqrt{\frac{12 \xi_\rho }{\lambda_\rho}}\,H \, \gtrsim\,  f_a \label{summaryStarobinsky} \;.
\end{align}
This result for $f_a^{\text{(inf)}}$ coincides with its counterpart in \eq \eqref{summaryNonminimal} derived in the Palatini formulation of GR with an identical lower bound on  $\xi_\rho$. Furthermore, the upper bound on $\xi_\rho$ is the same as \eq \eqref{boundPalatini}, derived in Palatini Higgs inflation (taking into account that $\Omega^2\approx N$). Thus, the additional kinetic mixing, which arises in metric GR (see \eq \eqref{actionEinstein}), does not play a role. 

\begin{figure*}
	\centering
	\includegraphics[width=1\textwidth]{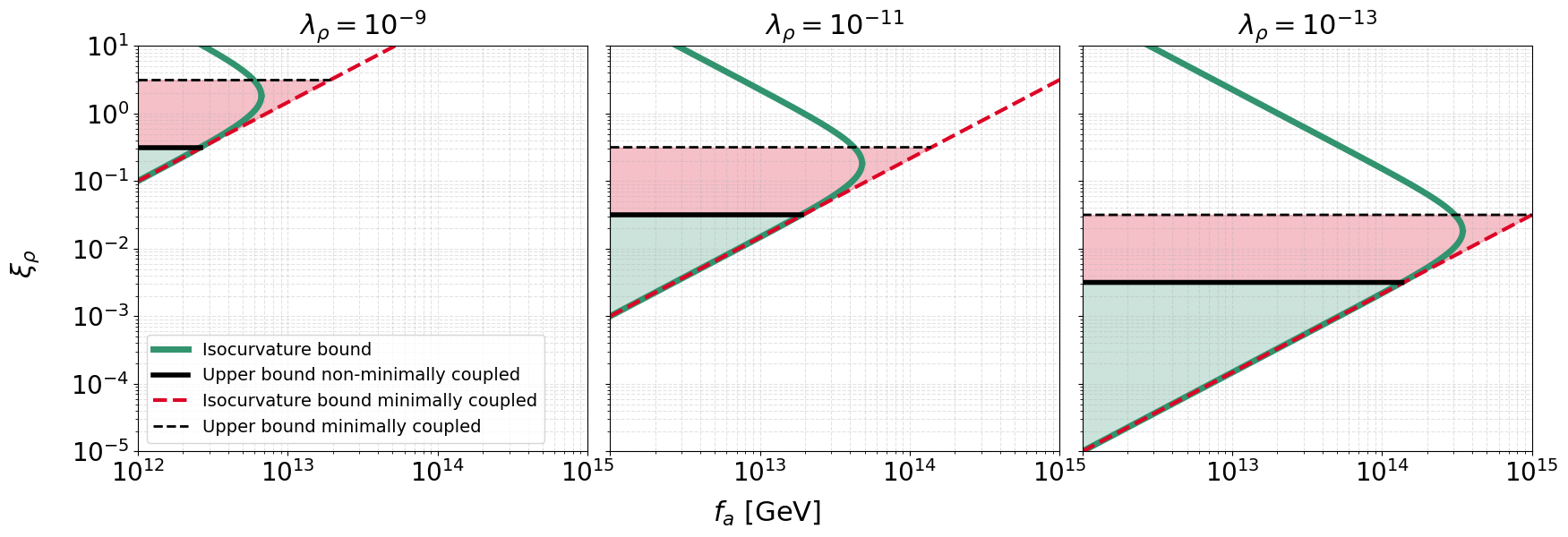}
	\caption{Constraints on $\xi_\rho$ in Starobinsky inflation as a function of $f_a$, for $H=10^{13}$ GeV and some choices of $\lambda_\rho$. The green line represents the lower bound on $\xi_\rho$ for which the inflationary decay constant \eqref{fInfStarobinskyNonminimal} fulfills the isocurvature bound \eqref{isocurvatureBound}. The red curve show the would-be lower bound on $\xi_\rho$ if the inflaton were minimally coupled, which is found by plugging
		\eq \eqref{faInfMinimalApprox} into \eq \eqref{isocurvatureBound}. Furthermore, the solid black line represents the upper bound on $\xi_\rho$ from imposing the non-backreaction condition of \eq \eqref{boundXiStarobinskyExact} (with $\Omega^2_0=N=50$) and the dashed black line corresponds to the would-be upper bound for the minimally coupled case. In the green region, isocurvature bounds are obeyed and the red region would only be viable if the influence of the inflaton non-minimal coupling on axions were neglected.}
	\label{fig:Starobinsky}
\end{figure*}

Plugging the largest admissible $\xi_\rho$ of \eq \eqref{summaryStarobinsky} into the isocurvature bound \eqref{isocurvatureBoundLifted}, we get 
\begin{equation} 
	\lambda_\rho 
	 < 10^{-7} \left(\frac{10^{12} \text{GeV}}{f_a}\right)^{7/3} \;.
\end{equation}
This result is very similar to its counterpart \eqref{boundLambdaPalatiniHI} in Palatini Higgs inflation. Thus, a very small $\lambda_\rho$ is required to alleviate isocurvature bounds.
Because of the condition $T_{\text{max}}<f_a$ of not restoring PQ symmetry during reheating, the precise value of the largest admissible $\lambda_\rho$ depends the transition from inflation to radiation dominated expansion.\footnote
{In Starobinsky inflation, reheating is not instantaneous but proceeds more slowly, where details depend on the precise coupling to matter fields (see \eg \cite{Tsujikawa:1999iv,Gorbunov:2010bn,Fu:2019qqe,Aoki:2022dzd,Dorsch:2024nan,Graham:2025iwx}). As a result, the maximal temperature $T_{\text{max}}$ does not saturate the bound \eqref{TMax}. 
On the other hand, $H \sim 10^{13}\, \text{GeV}$ is larger in the Starobinsky scenario as compared to Palatini Higgs inflation, and so the \rhs of \eqref{TMax} evaluates to a larger number. Therefore, we expect the resulting bounds on $\lambda_\rho$ to roughly be on the same order in both models.}
In \fig \ref{fig:Starobinsky}, we show  the influence of the inflationary decay constant \eqref{fInfStarobinskyNonminimal} on isocurvature bounds for different values of $\lambda_\rho$.

\section{Further constraints}
In this paper, we have focused on the avoidance of isocurvature constraints. Of course, this is only a necessary but not a sufficient condition for the phenomenological viability of a given scenario. We refer the reader to \cite{Graham:2025iwx} for a discussion of further constraints, among which we shall briefly discuss the following two.

A non-minimal coupling to gravity with a parameter $\xi_\rho>1$ lowers the cutoff scale $\Lambda$, beyond which perturbation theory breaks down, below the Planck scale. In metric GR, $\Lambda\sim M_P/\xi_\rho$ \cite{Burgess:2009ea,Barbon:2009ya}, and so the requirement $T_{\text{max}}<\Lambda$ can lead to a an upper bound on $\xi_\rho$ that is stronger than the conditions discussed thus far.\footnote
{In \cite{Graham:2025iwx}, the more restrictive condition $\sqrt{M_P H}\lesssim \Lambda$ was imposed. This coincides with our requirement only if reheating can be approximated as instantaneous, \ie the bound \eqref{TMax} on $T_{\text{max}}$ is saturated.}
In contrast, Palatini gravity leads to the significantly higher $\Lambda \sim M_P/\sqrt{\xi_\rho}$ \cite{Bauer:2010jg} (see also \cite{Karananas:2022byw}). Therefore, it follows from $H\ll \sqrt{\lambda_\rho} M_P/\xi_\rho$ (\cf \eqs \eqref{summaryMinimal}, \eqref{summaryNonminimal}) that $T_{\text{max}}<\sqrt{M_P H}<\lambda_\rho^{1/4} M_P/\sqrt{\xi_\rho}<\Lambda$, \ie the cutoff scale does not further constrain the viable parameter space in Palatini GR.

During reheating, the axionic decay constant relaxes from $f_a^{\text{(inf)}}$ to its late-time value $f_a$. This can lead to a non-thermal restoration of the PQ-symmetry \cite{Tkachev:1995md,Kasuya:1997ha,Tkachev:1998dc,Kasuya:1999hy} due to non-perturbative effects, and possibly also the formation of problematic topological defects \cite{Kawasaki:2013iha,Harigaya:2015hha,Kearney:2016vqw,Kobayashi:2016qld,Co:2017mop,Co:2020dya,Ballesteros:2021bee}. Without considering the effect of non-canonical kinetic terms, it was suggested that avoidance of PQ-restoration could lead to a strong bound $\xi_\rho \lesssim 10^2 \lambda_\rho f_a^2/H^2$, although the precise numerical value depends on the model \cite{Graham:2025iwx}. We expect, however, that non-canonical kinetic terms strongly influence the evolution of $\rho$ and $a$ during reheating, and so a detailed investigation of this phenomenon -- and reheating in general -- remains to be performed.

	\begin{table}
		\renewcommand{\arraystretch}{3.2}
		\setlength{\tabcolsep}{7pt}
		\begin{tabular}{|c|c|c|c|}
			\hline
			Model &  $\xi_{\rho\text{,min}}$        & $\xi_{\rho\text{,max}}$ & $f_a^{\text{(inf)}}$                                                              \\ \hline
			$\xi_\sigma=0$ &  \multirow{4}{*}{$\dfrac{\lambda_\rho f_a^2}{H^2}$ }  &  $\sqrt{\lambda_\rho} \dfrac{M_P}{H} $                     & \multirow{4}{*}{$\sqrt{\dfrac{12 \xi_\rho }{\lambda_\rho}}H$} \\ \cline{1-1} \cline{3-3}
			Palatini        &       &   \textcolor{gray}{ $  \sqrt{\lambda_\rho} \dfrac{M_P}{H} $    }           *                          &                                                                                   \\ \cline{1-1} \cline{3-3}
			Palatini Higgs   &                                  &            \multirow{2}{*}{$ \dfrac{\sqrt{\lambda_\rho}}{\Omega_0} \dfrac{M_P}{H}$}                   &                                                                                   \\ \cline{1-1} 
			Metric Starobinsky        &                                  &                   &                                                                                   \\ \hline
		\end{tabular}	
		\caption{Summary of bounds on $\xi_\rho$ for different models. The lower bound comes from requiring $f_a^{\text{(inf)}}>f_a$. The upper bound is necessary to ensure that backreaction on the inflaton is avoided. *Note that the upper bound for Palatini inflation is necessary but not sufficient, and once we specify the potential we may get a stricter upper bound, as shown for Palatini Higgs inflation.}
		\label{tab_summary}
	\end{table}

\section{Conclusion}
For scalar fields, the presence of a non-minimal coupling to gravity is arguably more natural than its absence, and such couplings play a central role in many successful inflationary models. However, a non-minimal coupling $\xi_\sigma$ of the inflaton to gravity inevitably decreases the inflationary value $f_a^{\text{(inf)}}$ of the axion decay constant and thus worsens the compatibility with isocurvature bounds \cite{Rigouzzo:2025hza}. Conversely, for a minimally coupled inflaton, a non-minimal coupling $\xi_\rho$ of the radial PQ field can increase $f_a^{\text{(inf)}}$ and thereby relax isocurvature constraints \cite{Graham:2025iwx}. In this paper, we have combined these two effects and identified the conditions under which each of them dominates.

We have shown that $\xi_\sigma$ reduces the maximal viable value of $\xi_\rho$, primarily due to the requirement of not significantly modifying the derivative of the inflationary potential. This in turn suppresses the maximal enhancement of the inflationary decay constant $f_a^{\text{(inf)}}/f_a$ and therefore reduces the parameter region in which isocurvature bounds can be alleviated. Remarkably, whenever $\xi_\rho$ succeeds in alleviating isocurvature constraints, the same expression \eqref{previousResult} for $f_a^{\text{(inf)}}$ as in the  case of a minimally coupled inflaton still holds (\cf \tab \ref{tab_summary}). Furthermore, these findings are largely insensitive to the formulation of GR, although some quantitative features in metric GR remain model dependent.

We have explored these effects in two concrete inflationary scenarios -- Palatini Higgs inflation and Starobinsky inflation-- with all our main findings summarized in \tab \ref{tab_summary}. In both cases, a non-minimal coupling $\xi_\rho \sim 10^{-1}$ can alleviate isocurvature bounds, albeit at the cost of requiring a small self-coupling $\lambda_\rho \lesssim 10^{-9}$. It would be very interesting to perform a comprehensive phenomenological parameter scan of these models, in analogy to the analysis of \cite{Graham:2025iwx}. Furthermore, understanding the role of non-minimal couplings (and the resulting non-canonical kinetic terms) during reheating is an important next step, as it could significantly impact the relevance of non-perturbative effects.

\begin{acknowledgments}

\paragraph*{Acknowledgments} We thank Georgios Karananas and Misha Shaposhnikov for insightful feedback on the manuscript. C.R.~acknowledges support from the
Science and Technology Facilities Council (STFC). The work of S.Z.~was supported by the European Research Council Gravites Horizon Grant AO number: 850 173-6.

\textbf{Disclaimer:} Funded by the European Union. Views and opinions expressed are however those of
the authors only and do not necessarily reflect those of the European Union or European Research
Council. Neither the European Union nor the granting authority can be held responsible for them.
\end{acknowledgments}

\bibliography{Refs}

\end{document}